# Borobudur was Built Algorithmically

Hokky Situngkir
[hokky.situngkir@surya.ac.id]
Dept. Computational Sociology, Bandung Fe Institute
Center for Complexity, Surya University

**Abstract**

The self-similarity of Indonesian Borobudur Temple is observed through the dimensionality of stupa that is hypothetically closely related to whole architectural body. Fractal dimension is calculated by using the cube counting method and found that the dimension is 2.325, which is laid between the two-dimensional plane and three dimensional space. The applied fractal geometry and self-similarity of the building is emerged as the building process implement the metric rules, since there is no universal metric standard known in ancient traditional Javanese culture thus the architecture is not based on final master plan. The paper also proposes how the hypothetical algorithmic architecture might be applied computationally in order to see some experimental generations of similar building. The paper ends with some conjectures for further challenge and insights related to fractal geometry in Javanese traditional cultural heritages.

**Keywords:** Indonesia, Borobudur, fractal geometry, self-similarity, cube-counting method.

*…the beauty and delicate execution of the separate portions, the symmetry and regularity of the whole, the great number and interesting character of the statues and reliefs with which they are ornamented, excite our wonder that they were not earlier examined, sketched and described…*
-Thomas Stanford Raffles on Borobudur (History of Java)

## 1. Introduction

As a a legacy from the greatness of the past, there have been still a lot of mysteries behind the structures of Indonesian Borobudur Temple. Some of them are described eloquently in Miksic (1990: 44-46). The hypothetical propositions backed by science are still a few, especially when it is related to mathematical one. Yet, Borobudur has been worldly recognized as one of biggest wonders in human civilizations. The Borobudur was a built in the theological tradition from 760 to 825 AD Mahayana Buddhist, located in Magelang, Central Java, Indonesia. Glance view of the Borobudur brings us to see the complexity of architectural design implemented to the temple with specific and unique appearance relative to other architectural and historical wonders, e.g.: Egyptian and Mayan Pyramid, Cambodian Angkor Wat.

The temple is built upon 123 x 123 $m^2$ land and comprises 6 square platforms and 3 circular platforms on top with a dome as the highest points. The decoration of the temple presents 2,672 detail relief panels narrating Buddhist mythologies. There are 504 Buddha statues in Borobudur and various stupas, the Buddhism related mound-like and bell-shaped structure. At the circular platform of the temple, there are 72 Buddha statues seated inside perforated stupa. A description related to history of reconstruction, site description, anthropological and archaeological perspective of the site are elaborated in Soekmono (1976) as the temple is closely related to Indonesian social living, even at the modern times (Vickers, 2005). It is also worth to note a good introduction the functional part of temples, in general , in Indonesian culture as described in Soekmono (2005). The late traditional kingdoms in Indonesian archipelago inherited various temples, and Borobudur is one of the greatest.

In the other hand, by the end of the previous millennium, a lot of works and researches have shown conjectures not to approach traditional culture by using conventional geometry. The geometry of fractal (Mandelbrot, 1983) has opened the door to see traditional cultures in the fractal perspective. The work of Elgash (1999), for instance, discusses how traditional ethnic groups in Africa build the fractal structure of architectures and in other crafts. The work of Wolfram (2002) has shown some alternating point of view on traditional of Eastern cultures and recently even put foundations to algorithmic architectural studies by incorporating cellular automata. In Indonesia, explanation of fractal geometry on traditional motif of fabric, batik, has even brought to implementation of generative art of batik (Situngkir, 2008). As related to the complexity studies, fractal geometry have provided us with a way reading the complexity emanated from aspects of Indonesian traditional culture – as to the archipelago is one of the richest place with diverse ethnicities (*cf.* Situngkir, 2005).

The mathematical study for Borobudur's architectural design has once related to answer the question about the metric system used by ancient Javanese to build such giant buildings with good measurement. While the anthropological revealed that Javanese used *tala* system (metric system with length measurement defined as the length of a human face from the forehead's hairline to the tip of the chin or the distance from the tip of the thumb to the tip of the middle finger when both fingers are stretched at their maximum distance), the survey as elaborated in Atmadi (1988) showed there is a ratio used between parts of Borobudur. There is part of Head : Body: Foot (9 : 6 : 4) that is met in horizontal and vertical measurement of the temple. This is shown in figure 2. Furthermore, this study is related to Buddhism cosmology as shown in Long & Voute (2008). The latest showed how the temple is not only related to religious meaning, but also ancient astronomy.

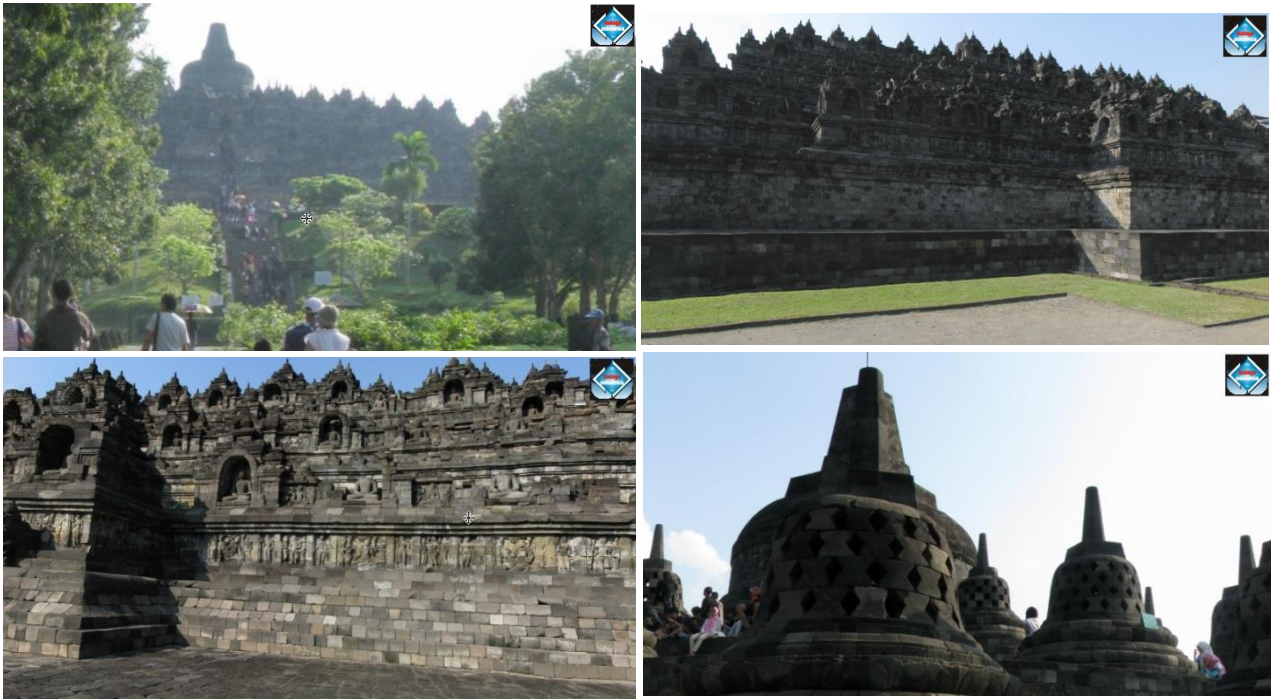

**Figure 1.** Borobudur

Glance observation to Borobudur, we can see some apparent aspects of self-similarity – a foundation of the fractal geometry. As the shape of stupa is presented in any view as main elements of the temple, some observers sometimes see the Borobudur itself as a giant stupa. The paper is motivated by the incorporation of fractal geometry to see the architectural design that would expect more progress on revealing some interesting facts of Borobudur related to fractal geometry. The paper is constructed as follows. The next section discusses some aspects of self-similarity and fractal geometry in the Borobudur architectural design. This is followed by some discussions on the geometry of stupa as the elements of the temple. The paper ends with some further works and concluding remarks.

**2. Fractal Geometry of Borobudur**
French scholar, Paul Mus, stated that the conical form of stupa is reflected in Borobudur in double expression. He said that the temple is an open-flat stupa, but as soon as people stay inside it, the temple expressed the idea of a closed world (*c.f.:* Moens, 1951). This view is obvious as we see figure 2 showing the vertical and horizontal cut of the temple as a whole. The silhouette of the temple is more like a stupa of which becomes the basic element comprising the whole construction. Thus, Borobudur can be seen as an 3 dimensional object of stupas within stupas. A quick survey to the site, we could discover a lot of construction elements showing the cone-like stupas in various sizes. There have been a lot of interpretations of what symbolized as stupa are (*c.f.*: Govinda, 1976). However, one of the most popular among Buddhists are that stupa symbolizes the enlightened mind of Buddha. This is what might want to be presented in the buildings of Borobudur as a giant stupa.

In Borobudur, there are several sizes of stupa-like-shaped symbols in various measurements. At the top of the building, there are 72 stupas within Buddha's relic inside. However, in the lower square floors (namely the *Rupadhatu* and *Kamadhatu*), stupa-like elements are shown as ornaments above the long balustrade decorated with reliefs telling the story of Buddha. In fact, we can also seen a the Buddha statues (there are 104 in the first level, followed by 104 more in the second level, 88 in the third, and 54 at the top of square-shaped floors) covered by bell-shape forms as stupa. A measurement of all apparent forms of stupas is shown in figure 3.

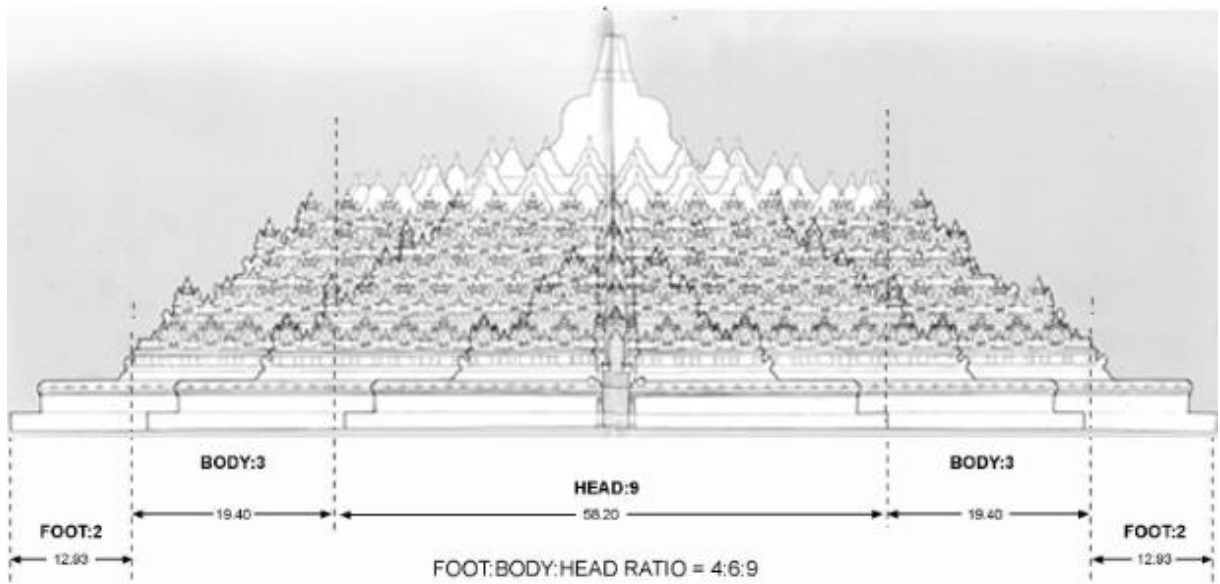

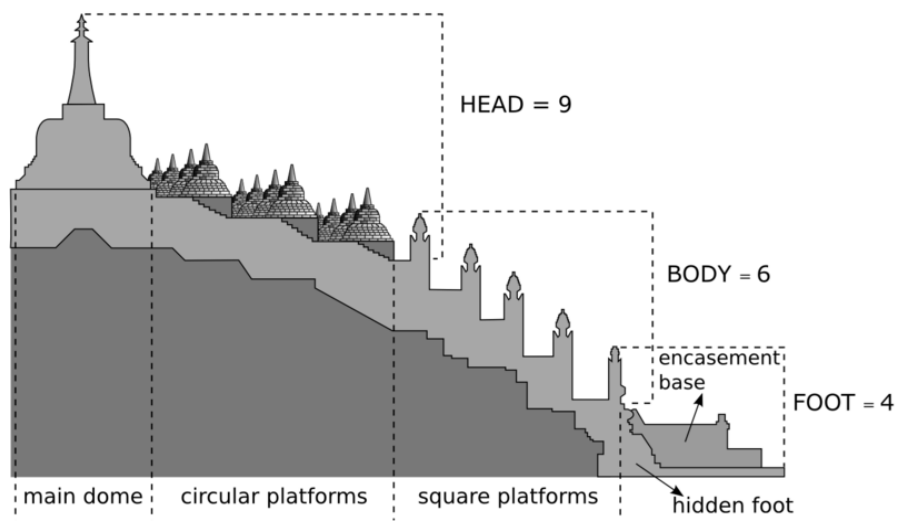

**Figure 2.** The Head-Body-Foot ratio in Borobudur vertical (*below*) and horizontal (*top*) wise

Interestingly, the log-log plot shown the number of the stupas (from one Borobudur treated as stupa to the tiny ornaments within the temple) as the function of the measurement (width/horizontally and height/vertically) draws a straight line as power function. Thus the measurement of the stupa within our measurement show the relation,

$$N(s_i) \sim s_i^{\delta} \qquad (1)$$

between the number of stupa $N(s_i)$ as function of width and height of the stupa for $i$ denotes the level of measurement and thus $\delta$ the power of the relation.

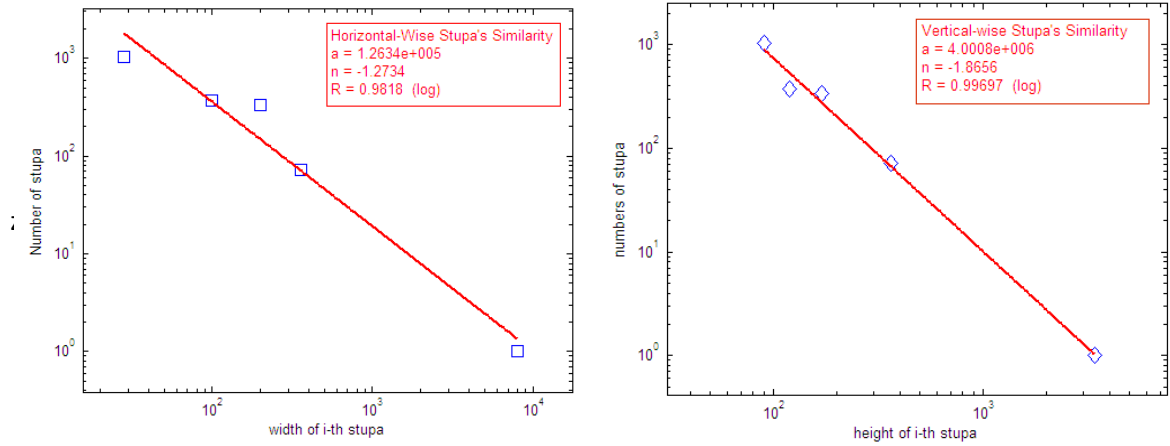

**Figure 3.** Self similarity of stupas in various scales within Borobudur

Furthermore, the challenge is to calculate the fractal dimension of the Borobudur temple as a whole building. In order to do this, we use the cub-counting method for the Minkowski–Bouligand dimension or box-counting dimension (Falconer, 2003:41-8 & Barnsley, 1988:172-95). Since the architectural model is always in 3-dimensional shape, if we denote Borobudur as the three dimensional $A \in F(X)$ where $(X, \varepsilon)$ is the metric space, we define that for $\forall \varepsilon > 0$ we have $N(A, \varepsilon)$ as the smallest number of cubes with the length of side is $\varepsilon > 0$ to cover the whole $A$. Here, Borobudur, denoted as $A$ is defined to have fractal dimension,

$$D = \lim_{\delta \to 0} \left( \frac{\ln(N(A, \varepsilon))}{\ln(1/\varepsilon)} \right) \qquad (2)$$

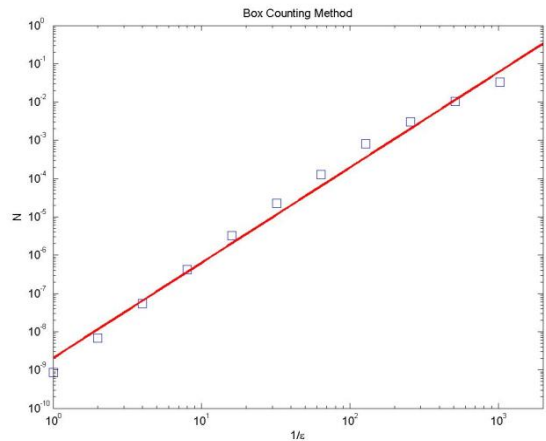

**Figure 4.** Box Count Method calculates the fractal dimension of the Borobudur.

Practically, the calculation is brought by using counting the number of cubes with side length $\left( 1/2^n \right)$ that intersect the building, to have $N_n(A)$. Thus, we have the fractal dimension to be

$$D = \lim_{n \to \infty} \left( \frac{\ln(N_n(A))}{\ln(2^n)} \right) \quad (3)$$

The cube counting is shown in figure 4 depicting that we have the Borobudur fractal dimension to be

$$D_{borobudur} = 2.3252. \quad (4)$$

As noted in Mandelbrot (1983: 468), a fractal dimension $2 < D < 3$ characterizes a fractally fragmented 3-dimensional object. The value also reflects that Borobudur is more likely to be experienced as two-dimensional object than a huge cube covering the whole temple in a single count of huge cube. Thus Borobudur is not a cone, even though we perceived it as cone-like shape in general, and stupa, as the basic elements of it, is too. The latest is discussed in the next section.

**3. Algorithmic Architecture Hypothesis**
As it has been discussed in the previous section, the traditional Javanese metric system (*tala*) can vary from person to person. While Borobudur must be built by incorporating lots of workers, the architect (named Gunadharma, but we do not know a lot about him), must apply a rule in order to build such mega-structures did not turn into a fiasco. Elaboration from the proposal of Atmadi (1988) on the ratio 4:6:9 employed in Borobudur architecture, we can hypothetically made a computational experiment using algorithmic rule in which the temple is built from the beginning – and in advance, see it's relation with the form of stupa as the element of building.

We propose the rule of the construction, i.e.: placing stones next to and above on others iteratively, horizontally and vertically. Placing stones side by side, the length of the next level of sequence should be made as reducing one third of the previous level, while the height should be as adding half of the previous one. Mathematically, we can write,

$$x(l+1) = x(l) - \frac{x(l)}{3} \quad (5)$$

$$y(l+1) = y(l) + \frac{y(l)}{2} \quad (6)$$

where $x(l)$ is the length and $y(l)$ is the height at level $l$. It is easy to understand that the diminishing factor horizontally is $r_{horizontal} = 2/3$ and the inverse, $r_{vertical} = 1\,1/2$, the growing factor vertically. Should the micro-rule of the temple following such steps, the ratio as conjectured by Atmadi (1988) would be met.

We do computational experiment to do such algorithm as described in eq. (5) and (6) and by changing the width of the initial level we could adjust the result with the one available to be observed in Borobudur. When it comes to different shape of initial level (the top circle levels and the square ones in lower level), we can also adjust the shape of the initial form, be it cube or a cylinder. The experiment is presented in figure 5. It is interesting to find that some forms are well suited with those we can see at the large scale of Borobudur. Nonetheless, the smaller ornamentation is crafted in such away possibly also use such rules, e.g.: making the small stupa in the lower levels as well as the perforated stupa at the above.

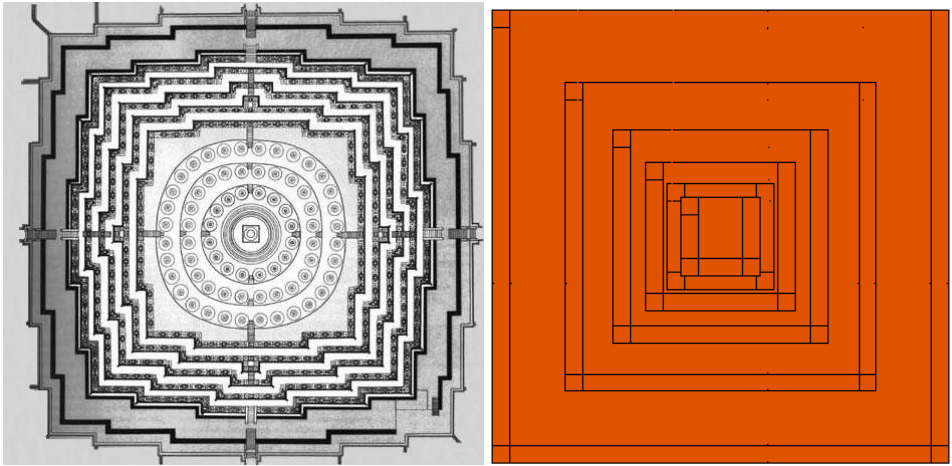
*Top view square lower levels*

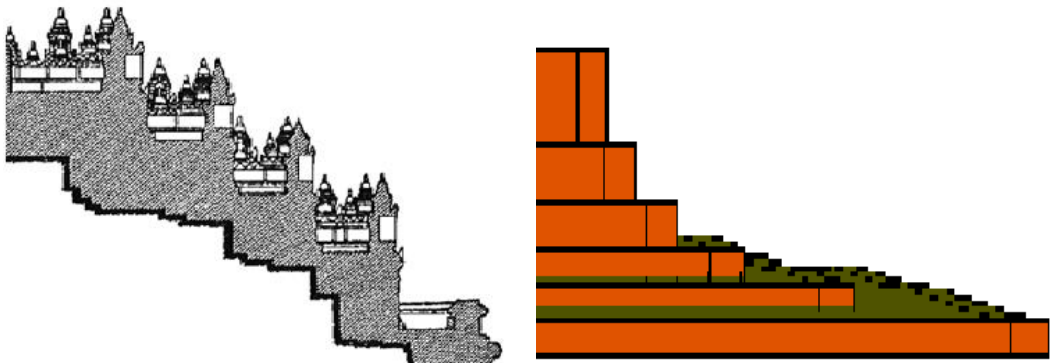
*Side view square lower levels*

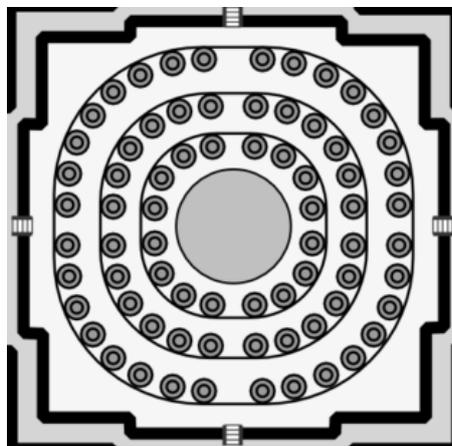
*Top view circle upper levels*

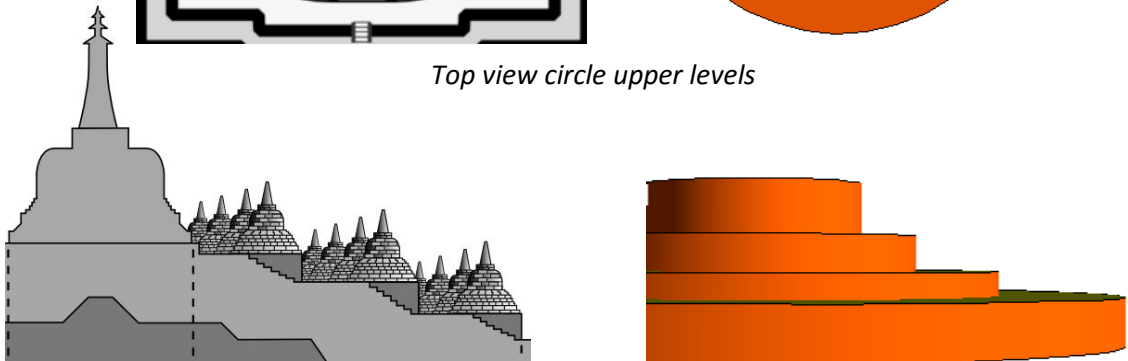
*Side view circle upper levels*

**Figure 5.** Elaborated version of the 4:6:9 hypothesis to algorithmically generate parts of Borobudur (*left*) and generic one (*right*).

When we alternate the initial element to be a sphere, we found formation that looks similar to the stupa, the hypothetical basic element of the temple. Some parts are having the same starting point to erect to the top. It is worth to note that this demonstration is not related directly to some theological or religious aspects of which stupa is frequently symbolized. However, the curiosity might be expanded to some issues relating the structure of stupa and it's its symbolization with the interesting geometry that is shown here.

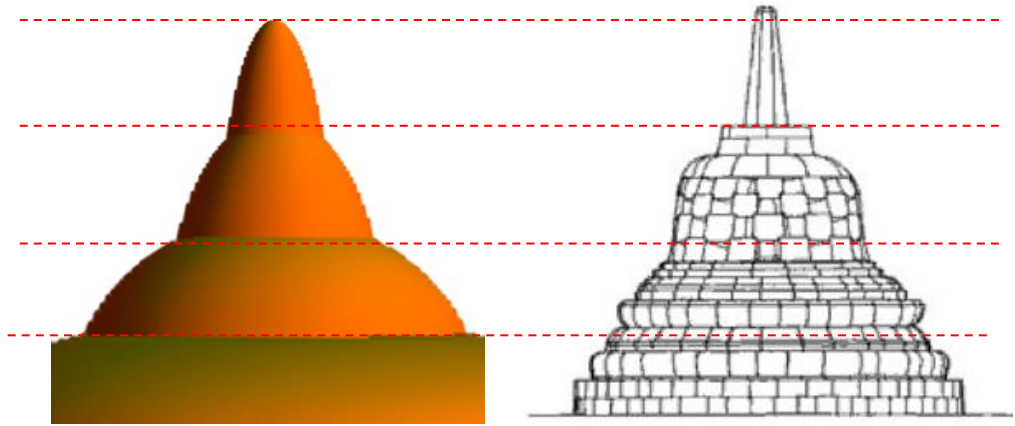

**Figure 6.** Hypothetical Algorithmic Generated Borobudur's Stupa (*left*) compared to the real one (*right*).

**4. Concluding Remarks & Further Works**
Algorithmic way that was incorporated in constructing Borobudur's architecture is a strong possibility for some issues related to the lack of standard metric system attached to ancient Javanese society and the closeness of Javanese culture with the fractal geometry that also found in traditional fabric, batik. Thus, we can say that while the inspiration of the building of Borobudur temple is religious issue, i.e.: Buddhism, the architecture is more likely strongly connected to the ancient Javanese culture. Borobudur temple was built as building a single and small stupa, but the way to making it was incorporated the technique of self-similarity. However, the emerged construction is eventually a kind of algorithmic fractal mega-architecture. The complexity of Borobudur is emerged from simple rules of building stupa as the fractal geometry applies.

The calculated fractal dimension of Borobudur is *2.325*, a number that shows the realm of the structure that is in between the two dimensional form and the three dimensional conic (or bell) shaped construction. This shows how self-similarity does exist and it is a theoretical challenge for interdisciplinary works among geometry, statistical analysis, computer sciences, anthropology, archaeology as well as mechanics to reveal deeper insights related to the dimension calculated. While in the previous works (Situngkir, 2008) the discussions have brought us to the interesting facts related to tradition fabric that also emanated applied fractal geometry, more observation and analysis related to the fractal aspects in cultural heritage might be appealing.

**Acknowledgement**
Author thanks Surya Research International for support in which period the paper is written and BFI colleagues for discussions on the rough draft of the paper.